\documentclass{nature}

\usepackage{mathrsfs,amsmath}
\usepackage{graphicx}
\usepackage{dcolumn}
\usepackage{bm}
\usepackage{amsmath}
\usepackage{color}
\usepackage{epstopdf}
\usepackage{textcomp}
\usepackage{cite}
\usepackage{verbatim}
\usepackage{gensymb}
\usepackage[dvipsnames]{xcolor}
\DeclareGraphicsExtensions{.pdf,.eps,.png,.jpg,.mps} 
\usepackage [english]{babel}
\usepackage [autostyle, english = american]{csquotes}
\MakeOuterQuote{"}

\begin{document}
\title{Femtosecond X-ray Fourier holography imaging of free-flying nanoparticles}
\maketitle
\author{Tais Gorkhover$^{1,2,3}$, Anatoli Ulmer$^{1}$, Ken Ferguson$^{2,3}$, Max Bucher$^{1,2,4}$, Filipe Maia$^{5}$, Johan Bielecki$^{5,6}$, Tomas Ekeberg$^{7}$, Max F. Hantke$^{5}$, Benedikt J. Daurer$^{5}$, Carl Nettelblad $^{5,8}$, Jakob Andreasson$^{5,9,10}$, Anton Barty$^{7}$, Petr Bruza$^{9}$, Sebastian Carron$^{2}$, Dirk Hasse$^{5}$, Jacek Krzywinski$^{2}$, Daniel S. D. Larsson$^{5}$, Andrew Morgan$^{7}$, Kerstin M\"{u}hlig$^{5}$, Maria M\"{u}ller$^{1}$, Kenta Okamoto$^{5}$, Alberto Pietrini$^{5}$, Daniela Rupp$^{1}$, Mario Sauppe$^{1}$, Gijs van der Schot$^{5}$, Marvin Seibert$^{5}$, Jonas A. Sellberg$^{5,11}$, Martin Svenda$^{5}$, Michelle Swiggers$^{2}$, Nicusor Timneanu$^{5,12}$, Daniel Westphal$^{5}$, Garth Williams$^{2,13}$, Alessandro Zani$^{5}$, Henry N. Chapman $^{7}$, Gyula Faigel$^{14}$, Thomas M\"{o}ller$^{1}$, Janos Hajdu $^{5,6,9}$ \& Christoph Bostedt$^{2,3,4,15}$}

\begin{affiliations}

\item Institut f\"{u}r Optik und Atomare Physik, Technische Universit\"{a}t Berlin, Hardenbergstr. 36, 10623 Berlin, Germany
\item Linac Coherent Light Source, SLAC National Accelerator Laboratory, Stanford, California 94309, USA
\item PULSE Institute and SLAC National Accelerator Laboratory, 2575 Sand Hill Road, Menlo Park, CA 94025, USA.
\item Argonne National Laboratory, 9700 South Cass Avenue, Lemont, IL 60439, USA
\item Laboratory of Molecular Biophysics, Department of Cell and Molecular Biology, Uppsala University, Husargatan 3 (Box 596), SE-751 24 Uppsala, Sweden
\item European XFEL GmbH, Holzkoppel 4, 22869 Schenefeld, Germany
\item Center for Free-Electron Laser Science, DESY, Notkestrasse 85, 22607 Hamburg, Germany
\item Division of Scientific Computing, Department of Information Technology, Science for Life Laboratory, Uppsala University, L\"{a}gerhyddsv\"{a}gen 2 (Box 337), SE-751 05 Uppsala, Sweden
\item ELI Beamlines, Institute of Physics, Czech Academy of Science, Na Slovance 2, CZ-182 21 Prague, Czech Republic
\item Condensed Matter Physics, Department of Physics, Chalmers University of Technology, Gothenburg, Sweden
\item Biomedical and X-ray Physics, Department of Applied Physics, AlbaNova University Center, KTH Royal Institute of Technology, SE-106 91 Stockholm, Sweden
\item Department of Physics and Astronomy, Uppsala University, Box 516, SE-751 20 Uppsala, Sweden
\item NSLS-II, Brookhaven National Laboratory, PO BOX 5000, Upton, NY 11973, USA
\item Research Institute for Solid State Physics and Optics, 1525 Budapest, Hungary
\item Department of Physics, Northwestern University, 2145 Sheridan Road, Evanston, IL 60208, USA


\end{affiliations}
\date{\today}

\begin{abstract}
Ultrafast X-ray imaging provides high resolution information on individual fragile specimens such as aerosols\cite{Loh2012}, metastable particles\cite{Barke2015}, superfluid quantum systems\cite{Gomez2014} and live biospecimen\cite{VanderSchot2015}, which is inaccessible with conventional imaging techniques. Coherent X-ray diffractive imaging, however, suffers from intrinsic loss of phase, and therefore structure recovery is often complicated and not always uniquely-defined \cite{Seibert2011,VanderSchot2015}. Here, we introduce the method of in-flight holography, where we use nanoclusters as reference X-ray scatterers in order to encode relative phase information into diffraction patterns of a virus. The resulting hologram contains an unambiguous three-dimensional map of a virus and two nanoclusters with the highest lateral resolution so far achieved via single shot X-ray holography. Our approach unlocks the benefits of holography for ultrafast X-ray imaging of nanoscale, non-periodic systems and paves the way to direct observation of complex electron dynamics down to the attosecond time scale. 

\end{abstract} 

High-resolution imaging of single non-periodic nanoparticles remains a great challenge. Electron microscopy requires the samples to be frozen and deposited on a substrate, potentially modifying their structure and functionality. Optical imaging techniques are limited in resolution. In contrast, intense femtosecond X-ray pulses from free-electron laser (FEL) sources enable "diffraction-before-destruction"\cite{Neutze2000} coherent diffractive imaging (CDI) of individual nanospecimen within a single exposure. Here, a single specimen is injected into the FEL focus from its native environment at room temperature and X-ray diffraction patterns are recorded long before the sample is vaporized by the FEL pulse\cite{Seibert2011, Chapman2011,Bostedt2012, Gorkhover2016}.

The central problem of all indirect imaging methods, such as CDI, is the intrinsic loss of phase during the measurement. Thus, structure recovery from CDI images is usually performed through iterative phasing algorithms which require solving a non-convex high-dimensional minimization problem. Much effort has been invested into phase retrieval algorithm development\cite{Miao1998,Marchesini2003,Chapman2010,Seibert2011,VanderSchot2015}.
Despite significant progress, structure reconstruction remains computationally demanding and requires guidance for selecting the object support. Particularly, noisy experimental data with missing data regions pose a challenge\cite{Seibert2011} and sometimes lead to incompatible solutions\cite{VanderSchot2015}.

Holographic approaches overcome the phase problem by encoding the relative phase between a reference object\textquotesingle s and the sample\textquotesingle s exit waves\cite{Gabor1948, Goodman2005}. In X-ray Fourier transform holography (FTH), a unique solution for the structure of the sample can be obtained from a simple two dimensional Fourier transformation (2D FT)\cite{Eisebitt2004, Schlotter2006, Marchesini2008, Geilhufe2014, Guehrs2010}. The highly desirable combination of X-ray Fourier holography (FTH) and CDI of free nanoparticles has been discussed theoretically in the past\cite{Eisebitt2004, Marchesini2008, Shintake2008, Guehrs2010}. The experimental realization has not been achieved so far as it is virtually impossible to prepare a well known and easily controlled reference with suitable spatial separation for randomly injected samples.


Here, we demonstrate a proof-of-concept in-flight holography experiment, which in contrast to classical X-ray Fourier holography, does not require careful positioning or precise preparation of the reference. Instead, the reference scatterer is randomly placed, and its size and position are directly extracted from the diffraction image as schematically illustrated in Fig. 1. Diffraction from two equally-sized spheres located in the plane perpendicular to the laser beam (Fig. 1a) is similar to Young \textquotesingle s double slit experiment: the intensity distribution envelope in the reciprocal space reflects the size of the spheres and the fine intensity modulation mirrors the relative distance between the spheres. Changes of the size of one sphere lead to an overlap of two characteristic envelopes (Fig. 1b). An off-set along the laser direction results in a distinct curvature of the fine modulations (Fig. 1c). Vice versa, the patterns from the reciprocal space can be unambiguously translated into a real-space three dimensional positions map based on the relative distances between the spheres directly encoded into the image. In in-flight holography, the smaller sphere can be regarded as a source of a reference wave front and the large sphere can be replaced by any unknown sample at a certain unknown distance. Using the information from the three-dimensional map, the structure of the sample can be reconstructed via Fourier inversion of the diffraction pattern.

In our proof-of-principle experiment, we used almost spherical gas-phase Xe clusters with diameters 30-120 nm as references and injected Mimi viruses with 450 nm sized quasi-icosahedral capsids as samples\cite{Xiao2009}. The experiment was carried out at the Linac Coherent Light source (LCLS) inside the LAMP end station\cite{Ferguson2015}, the experimental setup is illustrated in Fig. 2. We focused 100-femtosecond long X-ray FEL pulses with 1 nm wavelength to peak power densities up to 10$^{17}$ W/cm$^2$, which are necessary for single particle coherent diffraction imaging. 

The bio-sample and the cluster jets were overlapped inside the FEL focus using the instantaneous feedback from ion signal and X-ray diffraction patterns. The holograms were recorded using a pnCCD detector placed 735 mm downstream from the interaction region and covering up to a momentum transfer of q=0.3 nm$^{-1}$. 

A hologram of a Mimi virus is displayed in Fig. 3a. The imaginary part of the Mimi virus image from a reconstruction based on a Fourier inversion of the hologram (discussed in detail in the next paragraph and in Methods) is displayed in Fig. 3b. The shape and the size of the Mimi virus are in perfect agreement with results from previous studies\cite{Xiao2009, Seibert2011, Ekeberg2015} demonstrating that in-flight Fourier holography provides a correct and unique real-space shape of the sample. The structural information was retrieved in only few steps without the need of complicated iterative procedures. Note that due to the missing data from the FEL passage hole in the detector, the reconstruction is low frequency filtered and dominated by edge contrast information akin to a dark field microscopic image.

The reconstruction of the Mimi virus is guided by the information directly extracted from the hologram based on the scheme described in Fig. 1. In the reciprocal space, the diffraction pattern shown in Fig. 3a exhibits the characteristic streak pattern due to the quasi-icosahedral shape of the virus. The reference Xe cluster with the strongest signal produces the centrosymmetrical intensity distribution envelope fitted and highlighted in Fig. 3a, upper right quadrant of the hologram. From this envelope, we calculated the cluster diameter d = 74 nm using the Guinier approximation\cite{Guinier1955}. The fine modulations (magnified in Fig. 3c) convey information about relative distances between the Mimi virus and the reference clusters. Two independent modulations emphasized in Fig. 3c indicate the presence of at least two reference clusters.

The features in the reciprocal space pattern shown in Fig. 3a and c can be directly translated into a real-space three-dimensional map via two steps. First, a real space 2D projection onto a plane perpendicular to the FEL is created from an inverse 2D Fourier transformation (FT) of the hologram as presented in Fig. 4a. The highlighted region contains real-space cross-correlation terms (labeled with 1-3) between objects contributing to the hologram. The cluster-cluster and Mimi-Mimi auto-correlation terms are concentrated around the center of the FT. The distance between the center of the FT and the cross-correlation terms is the relative distance between the cross-correlation partners in the plane perpendicular to the FEL propagation as indicated in Fig. 4a and b by arrows. 

The cross correlations in Fig. 4a appear out-of-focus due to longitudinal offsets between the nanoparticles. Thus, in a second step, the cross-correlation terms are refocused using angular spectrum propagation\cite{Goodman2005,Guehrs2010,Geilhufe2014,Bostedt2010} as illustrated in Fig. 4c. The longitudinal distances between the objects can be recovered from a series of propagations into different planes perpendicular to the laser direction. Each propagation consists of a 2D FT of the hologram from Fig. 3a multiplied by a varying free space propagator (see Methods). Similarly to refocusing with a microscope objective, the sharp boundaries of the cross correlation terms emerge when the selected propagator term meets the real-space distance between the two correlating particles along the FEL axis as shown in Fig. 4c. 

Both reconstruction steps combined create a three-dimensional map of relative positions between the two clusters and the Mimi virus inside the FEL focal volume, as illustrated in Fig. 4d. For the final reconstruction of the Mimi virus presented in Fig. 3b, we used the refocused cross-correlation shown in Fig. 4a. labeled 1. The quality of the reconstruction was further enhanced through deconvolution with the reference cluster size determined from the envelope of the diffraction pattern as demonstrated in Fig. 3a (see Methods).


We have shown that in-flight holography can provide high fidelity images of highly asymmetrical samples. We recovered images from deformed specimen which have a more complex structure than the ideal Mimi virus. For example, the hologram in Fig. 5 a was most likely recorded from a conglomerate of three samples, while the hologram shown in Fig. 5 b stems from debris with a rich substructure. The reconstruction indicates that our method can be used to access the inner structure of a sample. The lateral resolution in our experiment was limited by the photon flux, similarly to non-holographic studies\cite{Schropp2010, Neutze2000,Seibert2011,Hantke2014}. We achieved a lateral resolution below 20 nm as shown in Fig. 5 C and 260 nm longitudinal resolution using small reference clusters with diameters 30-40 nm (see Methods). The longitudinal resolution can be in principle further improved using softer X-rays\cite{Barke2015}.

In summary, our experiment is the first demonstration of high fidelity imaging based on Fourier holography on free nanosamples. 
Our method is noise robust, reliable and can be extended to a variety of references and samples. With recent advances in single shot imaging with table-top sized lasers\cite{Rupp2016coherent}, in-flight holography may reach laboratory scale experiments due to its simplicity. This opens a novel avenue to study air pollution, combustion, cloud formation and catalytic processes on the nanoscale at the single-particle level. One could envision for example direct imaging of aerosol nucleation and droplet formation in their native environment without ensemble averaging effects\cite{Loh2012}. In-flight holography can be also easily extended towards time-resolved studies of complex, ultrafast electron dynamics\cite{Zherebtsov2011, Bostedt2012,Gorkhover2012,Gorkhover2016}. 

\clearpage


\begin{figure}
	\includegraphics[scale=0.4]{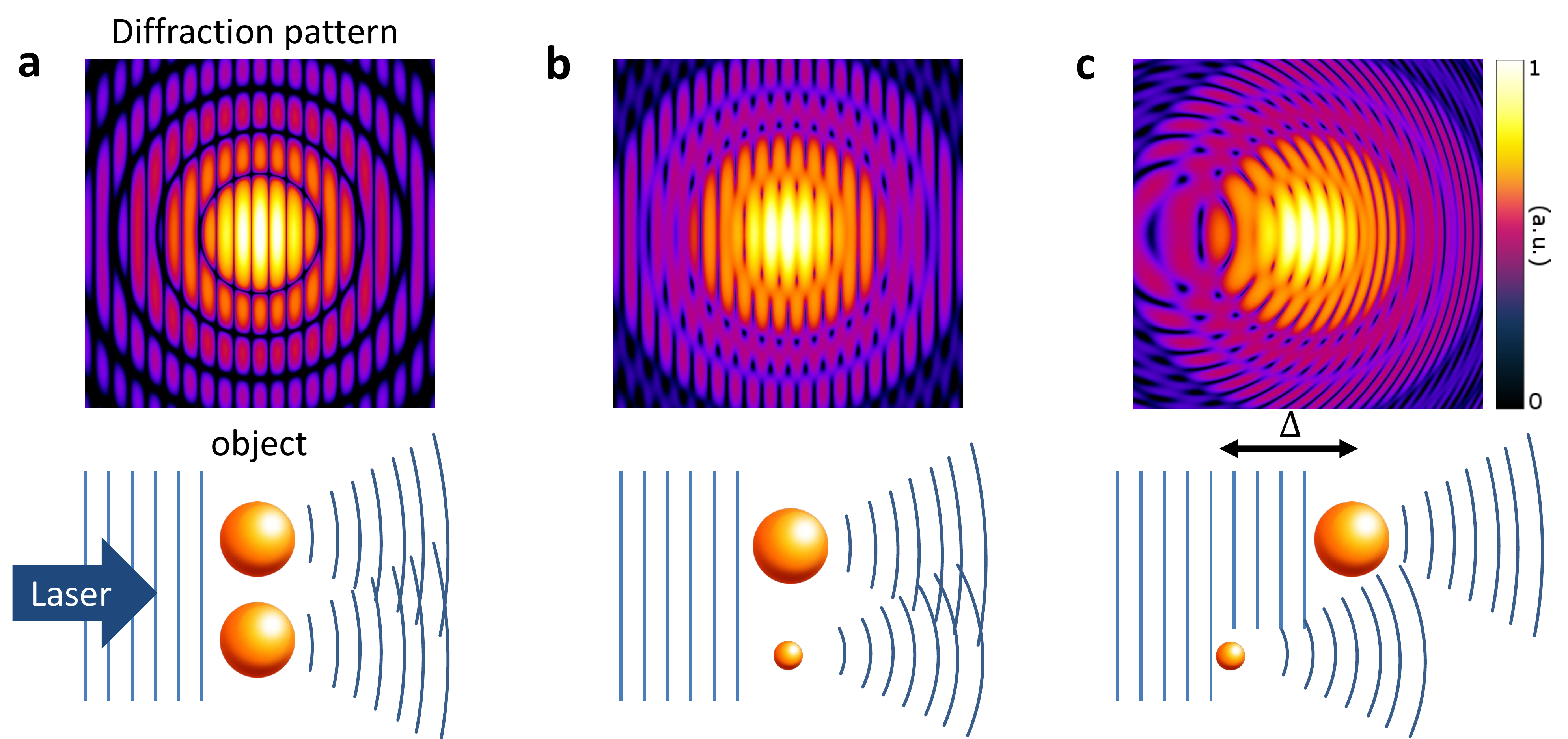}
	\caption{\textbf{
In-flight holography principle} Diffraction patterns in the first row are calculated from objects and geometries displayed in the second row.
	
	The diffraction pattern shown in \textbf{a} is produced by two equally sized spheres located in a plane perpendicular to the laser direction. It exhibits two dominant features. First, the ring-type envelope reflects the size of a single sphere. Second, fine straight modulation lines mirror the lateral distance between the spheres, similarly to Young\textquotesingle s double slit experiment. If one sphere is considerably smaller than the other, a second envelope with wider ring spacing appears as demonstrated in case \textbf{b}. Here, the different sizes of the spheres and the lateral distances are unambiguously encoded into the diffraction.
	In case \textbf{c}, the differently sized spheres are shifted along the laser direction.
	This shift in the real-space translates into curvature of the fine modulation lines in the reciprocal space. The combination of distinct diffraction features such as the envelopes and fine modulations carry a unique three-dimensional relative positions map of the two spheres. This map is used for structure recovery in in-flight holography, where the large sphere is replaced by an unknown sample. The smaller sphere can be regarded as a source of a Fourier holography-type reference wave front.}
	\label{fig:setup}
\end{figure}
\clearpage

\begin{figure}
	\includegraphics[scale=0.31]{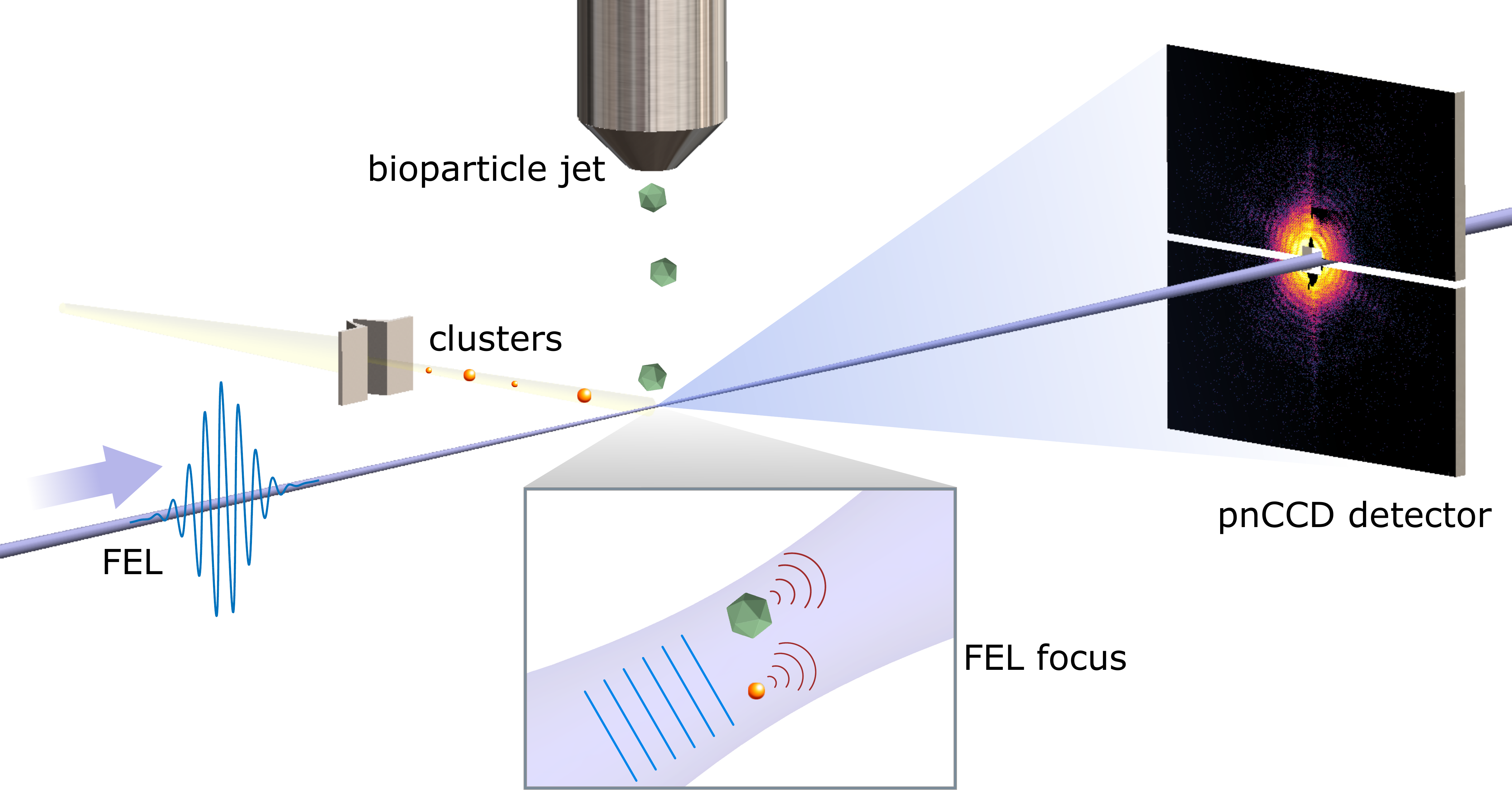}
	\caption{\textbf{Experimental setup.} The FEL beam is focused by a pair of Kirkpatrick-Baez mirrors in order to achieve FEL intensities necessary for single particle imaging. The bio particle jet and the Xe cluster jet are mounted perpendicular to each other and the FEL direction and overlapped inside the FEL focal volume. The X-ray photons diffract from randomly injected Xe clusters and nanosamples as schematically illustrated in the magnification panel. The resulting diffraction patterns are recorded on a pnCCD detector placed 735 mm downstream. The diffraction from the spherical clusters serves as a reference wavefront similar to a pinhole-like source.}
	\label{fig:setup}
\end{figure}
\clearpage

\begin{figure}
	\includegraphics[scale=0.26]{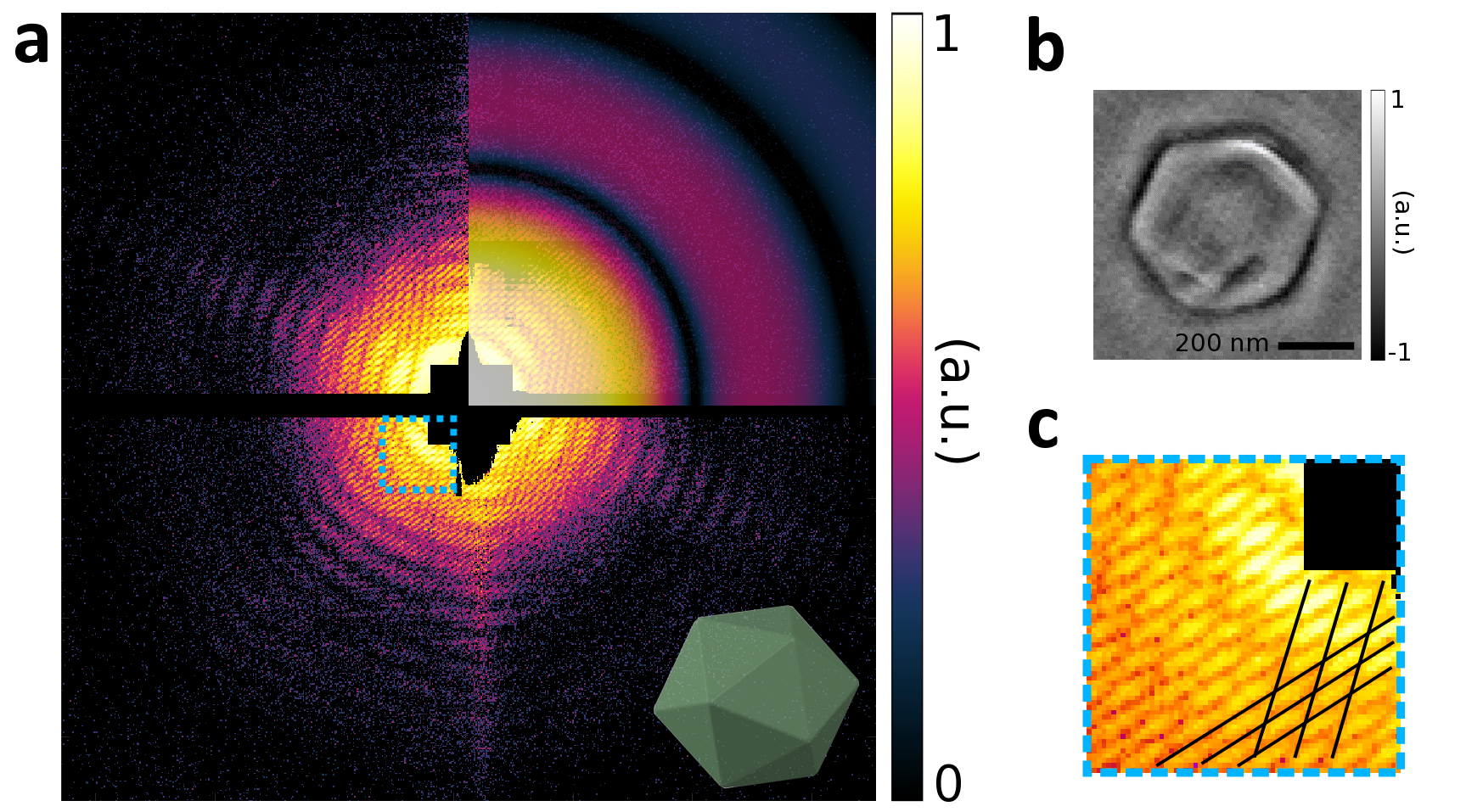}
	\caption{\textbf{Hologram and reconstructions}.  \textbf{a:} Hologram of a Mimi virus. The signature envelope from the reference cluster is visualized in the upper right quadrant. \textbf{b:} The reconstruction of the Mimi virus from the hologram corresponds to a 2D projection of a quasi-icosahedron as expected from previous experiments\cite{Xiao2009, Seibert2011, Ekeberg2015}. Note that the reconstruction is based on a two-step Fourier inversion without any preassumptions \textbf{c:} The magnified section of the hologram shown in \textbf{a} with contoured interference ripples. Two independent modulations indicate the presence of at least two reference scatterers.}
	\label{fig:results}
\end{figure}
\clearpage

\begin{figure}
	\includegraphics[scale=0.3]{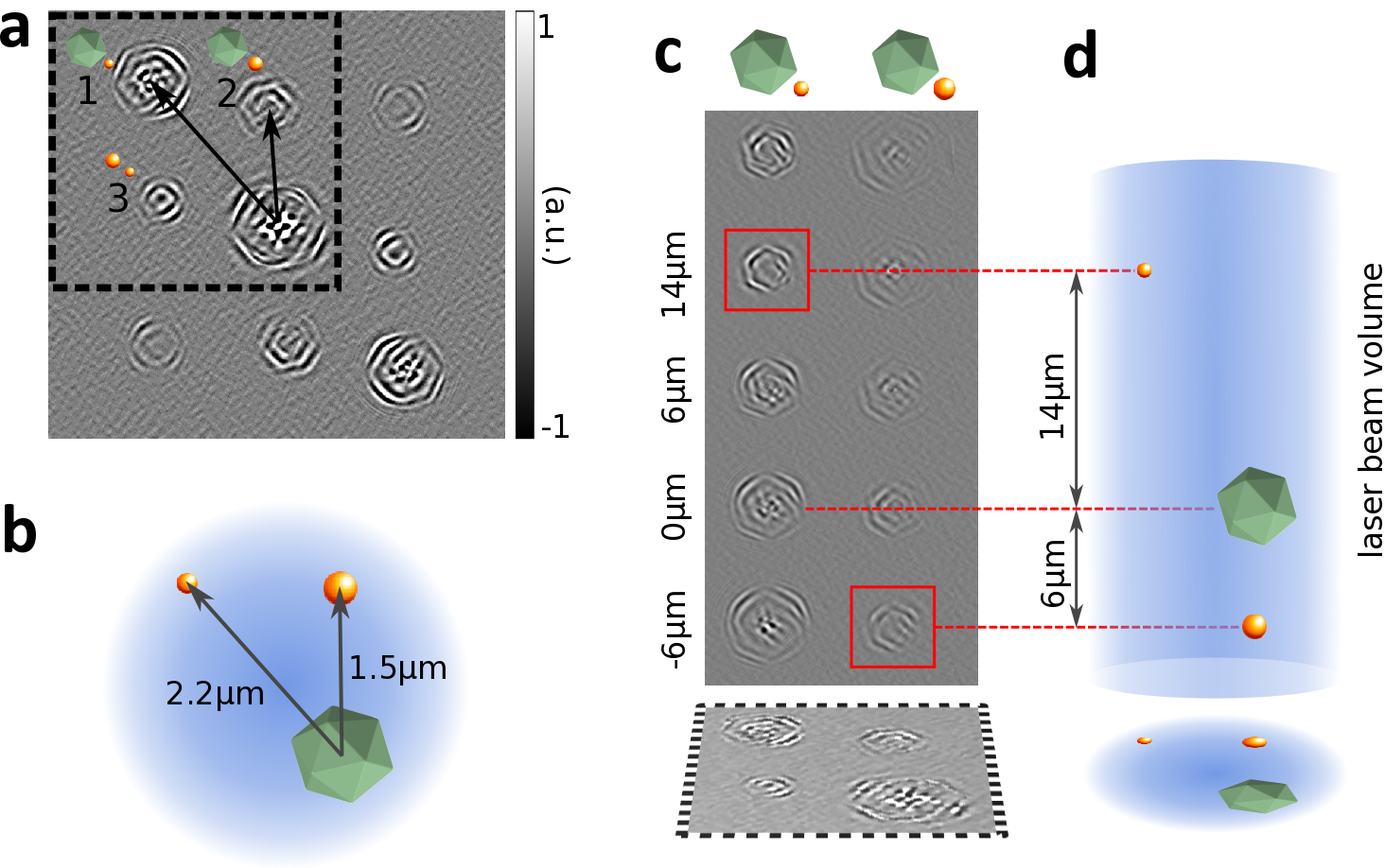}
	\caption{\textbf{Three dimensional reconstruction. }\textbf{a:} 2D Fast Fourier transformation (2D FFT) of the hologram displayed in Fig. 3a. The cross-correlation terms result from the interference between the spatially separated exit waves of the correlating particles and thus are displaced from the center. Cross-correlation 1 can be attributed to a Mimi virus and a cluster with d=74 nm, cross-correlation 2 to a Mimi virus and a cluster with d=100 nm, and cross-correlation 3 to clusters with d=74 and d=100 nm. The twin images of the cross-correlations are distributed point-symmetrically around the center. \textbf{b:} The 2D FFT shown in \textbf{a} contains positions of all three particles inside the plane perpendicular to the FEL. \textbf{c:} Using angular spectrum propagation, it is possible to recover the relative distances along the FEL axis. \textbf{d:} The combination of \textbf{a} and \textbf{b} creates a three-dimensional map of two clusters and the Mimi virus. The forth cross correlation is not discussed here as the corresponding cluster is too far from the Mimi virus. Note that there is no need for several references in order to reconstruct the Mimi virus.}
	\label{fig:simulation}
\end{figure}
\clearpage

\begin{figure}
	\includegraphics[scale=.6]{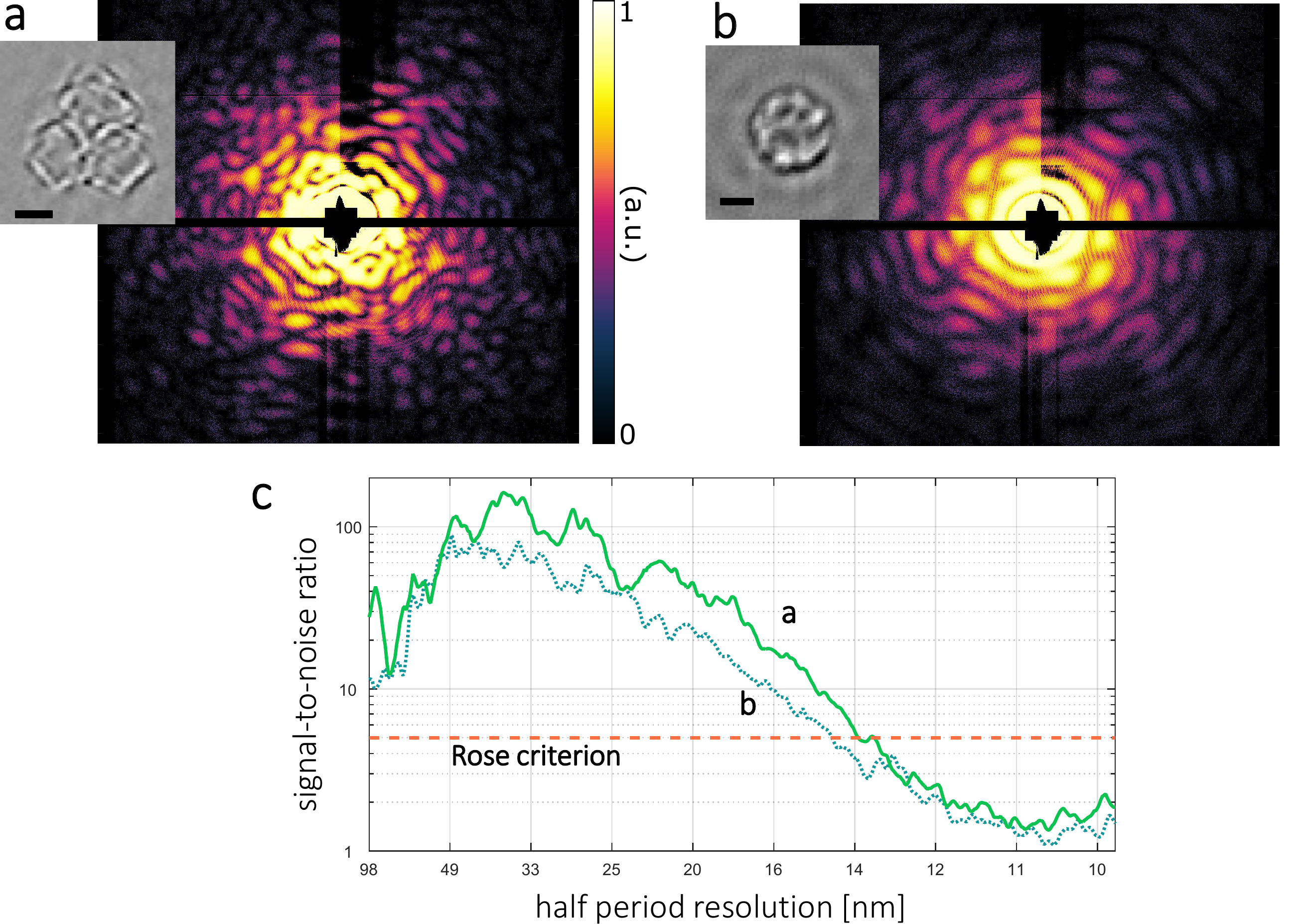}
	\caption{\textbf{Holograms of deformed specimen recorded with small reference clusters}. Panels \textbf{a} and \textbf{b} display holograms and their reconstructions as insets. The scale bar represents 100 nm. In \textbf{c}, the corresponding signal-to-noise ratio vs. half-period resolution is plotted. The central diffraction speckle from the reference cluster with diameter $\le$20 nm covers almost the entire detector area. }
	\label{fig:simulation}
\end{figure}
\clearpage

\begin{methods}

\subsection{Experimental Methods}
The experiments were performed at the LAMP instrument inside the AMO end station at the Linac Coherent Light Source (LCLS)\cite{Emma2010}. We used soft X-ray pulses with 1200 eV photons and pulse energies up to 2 mJ at 120 Hz repetition rate. The biological samples were injected in a solution through a gas dynamic virtual nozzle using a guiding helium gas flow\cite{Hantke2014}. The particle jet was focused down to 100 $\mu$m in diameter with an aerodynamic lens stack consisting of 6 apertures. The solvent evaporates during the injection. The Xe clusters were produced via a supersonic expansion through a conical nozzle with a diameter of 200 $\mu$m and an half opening angle of $4\degree$, which was attached to a commercial Parker Series 99 Pulsed Valve. The mean cluster size was varied between 30 and 120 nm in diameter by adjusting the backing pressure of the gas resovoir (from 7 bar to 30 bar) and the nozzle temperature (from 240 - 300 K). It has been demonstrated in previous experiments that within the achieved resolution, down to 8nm scale the large majority of Xe clusters are spherical\cite{Rupp2012,Gorkhover2012,Gorkhover2016}. The pulsed cluster jet was synchronized to the FEL at 10-30 Hz repetition rate with regard to the gas load faced by the pumps. The particle jet was skimmed down with a system of two fixed skimmers and an adjustable piezo skimmer reaching a cluster jet width between 100 $\mu$m and 3 mm.

The X-ray diffraction patterns were recorded via a pnCCD detector pair with 75 $\mu$m x 75 $\mu$m pixel size and 1024 x 512 pixels per detector half. The two detector parts were positioned 735 mm downstream from the interaction region.

\subsection{Requirements to coherence}
The longitudinal coherence of LCLS pulses in the soft X-ray regime has been estimated to 0.5 fs which translates to 0.17 $\mu$m in real-space\cite{Gutt2012}. In principle, forward scattering from two objects placed with an offset along the FEL beam always leads to a coherent superposition of the scattered wave fronts. Constructive interference at scattering angles greater than zero can occur when the path length difference between the wavefronts is smaller than the longitudinal coherence. In our case, interference at the outer edge of the detector can be expected from reference-sample pairs with maximum longitudinal distance of around 60 $\mu$m.

\subsection{Structure reconstruction}
The reconstruction from an X-ray hologram is sectioned in three steps and described below in greater detail. First, the images were calibrated using the dark noise level and masked due to detector imperfections. Second, the three dimensional composition between the sample and the reference was reconstructed. Third, the three dimensional information was used to refocus and refine through deconvolution. 

\paragraph*{{\textit{Image imperfections}}}
The intrinsic noise of the detector was substracted using dark calibration and common mode correction, followed by a correction for beamline background X-ray scattering and readout artifacts\cite{Strueder2010}.

A binary mask from missing, dead, highly fluctuating and saturated detector pixels was generated for each diffraction pattern. To suppress reconstruction ringing artifacts from an oscillating point spread function, the transition from masked areas to areas of data pixels was smoothed out. This was accomplished by first dilating the binary mask by 5px and afterwards applying a Gaussian filter with $\sigma = 5$px, to ensure a continuous transition to zero. We found that the such applied changes of the data values lead only to minor changes in the reconstruction contrast while ringing in the surrounding areas is efficiently suppressed.\\

\paragraph*{{\textit{Three dimensional reconstruction}}}
The samples and the reference clusters are almost transparent for the X-rays. Thus, first Born approximation and the small-angle approximation are valid. In the far field region, the scattering pattern $O$ equals the absolute square of the Fourier transform of the projection of the object's electron density $o$ along the FEL axis:

\begin{align}
    O(\mathbf{q_{\bot }}) = \left|\mathscr{F}[o(\mathbf{x_{\bot }})] \right|^2
\end{align}

Hereby $\mathbf{x_{\bot }}$ and $\mathbf{q_{\bot }}$ denote the coordinates perpendicular to the FEL beam in real space and the scattering vector in momentum space, respectively. By introducing a reference scatterer $r(x_{\bot }-x^0_{\bot })$, the exit waves of both particles interfere and the scattering pattern becomes a Fourier hologram. It is important to note that here the particles were smaller than the depth of field (DOF) and the exit wave of each particle can be interpreted as a projection of its electron density. At the same time the distance between the particles along the FEL axis can be larger than the DOF and therefore results in 3D information about the positions of the scatterers in regard to each other.

In the first step of the reconstruction, a single inverse Fourier transformation of the recorded two-dimensional hologram $H(\mathbf{q_{\bot }})$ can be regarded as a translation from momentum space coordinates to real space coordinates and results in the Patterson map $P(\mathbf{x_{\bot }})$:

\begin{align}
    P(\mathbf{x_{\bot }}) = \mathscr{F}^{-1} H(\mathbf{q_{\bot}}) &=  \mathscr{F}^{-1} \frac{4\pi^2}{\lambda^2 D^2} \left| \mathscr{F} (o(\mathbf{x_{\bot }}) + r(\mathbf{x_{\bot }-x_{\bot }^0})) \right|^2 \nonumber\\
    &= \frac{2\pi}{\lambda^2 D^2} \Big( o^*(-\mathbf{x_{\bot }}) \otimes o(\mathbf{x_{\bot }})
     + r^*(-\mathbf{x_{\bot }}+\mathbf{x_{\bot }^0}) \otimes r(\mathbf{x_{\bot }}-\mathbf{x_{\bot }^0}) \nonumber\\
    &\quad + o^*(-\mathbf{x_{\bot }}) \otimes r(\mathbf{x_{\bot }}-\mathbf{x_{\bot }^0})
     + r^*(-\mathbf{x_{\bot }}+\mathbf{x_{\bot }^0}) \otimes o(\mathbf{x_{\bot }}) \Big).
\end{align}

with X-ray photon wavelength $\lambda$ and detector distance $D$. Hereby $\otimes$ describes the convolution operator and the terms $\psi^*\otimes\psi$ and $\rho^*\otimes\rho$ are the autocorrelation functions without direct phase information. The spatially separated cross-correlation terms $\psi^*\otimes\rho$ and $\rho^*\otimes\psi$ contain relative phases between the reference and the sample wave fronts (Fig. 4). With some knowledge about the reference (in our case: size and spherical shape), these terms allow for unambiguous reconstruction of the sample exit wave and thus the sample's structure. The lateral distance of the cross-correlation partners has to be at least twice the diameter of the exit waves to avoid overlapping of the terms with autocorrelation function in the center and enable reconstruction.

In a second step, the cross-correlation terms were refocused using the free space propagator. The sample and the reference clusters were randomly injected into the FEL focal volume with an offset between the sample and the reference along the optical axis as demonstrated in Fig. 1 and 4. Consequently, the sample\textquotesingle s exit wave propagates over a distance $z$ before it interferes with the references exit wave or vice versa. Assuming free propagation along the FEL beam (z-direction) the sample´s structure can be recovered by propagating the exit wave to the plane of the sample using the angular spectrum propagation $\psi_R(x,y,z)$\cite{Goodman2005}:

\begin{align}\label{eq:propgl}
    \psi_R(x,y,z) &= \mathscr{F}^{-1}e^{i z \sqrt{k^2-q_x^2-q_y^2}} H(q_x,q_y)
\end{align}

with the wave field vector $\mathbf{k}=\frac{2 \pi}{\lambda}\frac{\mathbf{r}}{r}$, $\mathbf{r}=(x,y,z)$ the real space coordinates and $k=|\mathbf{k}|$.
The refocusing of the hologram can be done by eye similar to the refocusing of a light microscope image as illustrated in Fig. 4 or using autofocus techniques. If the phase shift matches the real-space distance between the reference and the sample, the sample´s structure emerges with high contrast and sharp features. All reconstructions with refocused cross correlation terms were sorted according to the size and analyzed by hand for reference-reference and sample-sample cross correlations.

\paragraph*{{\textit{Refining the image of the virus}}}
In the last steps, the diffraction patterns were centered based on the reduction of phase ramps in real space and deconvolved using a Wiener deconvolution filter for the cases, where the reference cluster size was well known from radial plots\cite{Howells2001, He2004}. The reconstruction represents the convolution of the reference and the sample, thus the resolution is often limited by the reference scatterers size. However, using the knowledge that the produced reference clusters are spherical in good approximation within the reached resolution, it is possible to resolve even smaller structures. In the current photon energy regime, the diffraction pattern of a spherical particle is well described within Guinier's approximation\cite{Guinier1955}. This allows for deconvolution of the reconstruction. Here the Wiener deconvolution filter was used:

\begin{equation}
    F_W (q) = \frac{1}{S(q)} \frac{| S(q) |^2 \cdot \mathrm{SNR}(q)}{1 + | S(q) |^2 \cdot \mathrm{SNR}(q)}
\end{equation}

$S(q)$ describes the estimate for the reference and SNR$(q)$ the signal-to-noise ratio. As the real ratio is not known, here the square root of the scattering pattern is used as the SNR approximation. This approximation is justified by the fact that the Poisson noise is the main noise contribution and the deconvolution quality depends rather on the scale than the exact SNR. Additionally Wiener deconvolution assumes signal independent noise, which is not the case here. As a result SNR$(q)$ is here more an adjustable parameter than the genuine noise.

\subsection{Resolution estimation}
We used the signal-to-noise ratio of the power spectrum of each individual reconstruction to estimate the half-period resolution based on the Rose criterion.

The resolution in Fourier holography is usually limited by several constraints. In the ideal, detector limited system the resolution is determined by the numerical aperture while in the perfect reference limited system it is given by the size of the reference scatterer. This limitation is not necessarily caused by the fact that the reconstruction describes the convolution of both particles. With the knowledge about the reference morphology it becomes possible to deconvolve the reconstruction and therefore resolve even smaller features\cite{He2004}. The limitation hereby is set by the fact that the signal is modulated by the spectrum of the reference which has zero crossings. Therefore even the deconvolved reconstruction has a drop in its spectrum which translates into the resolution limit of a perfect reference limited system. Nevertheless a deconvolution changes the transmission function within the boundaries of the central speckle of the reference and therefore obtains resolution before the first zero crossing. 

In our case, the resolution is limited by photon flux or in other words Poisson noise. Thus, we estimated our resolution from the signal-to-noise ratio in each reconstruction. In order to estimate the noise level, empty areas at different positions of the Patterson map were used.

\subsection{Hit finding}
Bright images were extracted according to the total number of lit pixels and a 2D FFT was calculated. Potential holograms were filtered based on the presence of cross correlation terms as described in\cite{Hantke2014}. Further quality assessment was performed by eye. Within the course of this experiment, the hit rate changed considerably. For example the Mimi virus hologram displayed in Fig. 2 and 3 was selected from a run recorded during 15 minutes at 30 Hz rate and containing in total 15 bright holograms of the Mimi virus. The holograms from Fig. 5 were extracted from a 10 minutes run recorded at 30 Hz and containing over 100 holograms. The hologram hit rate varied strongly depending on fine adjustments of both sources. We expect that the hit rate can be significantly improved based on recently described injector developments\cite{Hantke2014}.

\end{methods}

    
    \begin{addendum}
     \item We would like to thank Jan Geilhufe, Erik Guehrs, Andreas Schropp and Stefan Eisebitt for many helpful discussions. T.G. acknowledges a Peter Ewald fellowship from the Volkswagen Foundation. Parts of this research were carried out at the Linac Coherent Light Source (LCLS) at the SLAC National Accelerator Laboratory. LCLS is an Office of Science User Facility operated for the U.S. Department of Energy Office of Science by Stanford University. This work is supported by the U.S. Department of Energy, Office of Science, Office of Basic Energy Sciences, Division of Chemical, Geological, and Biological Sciences, under contract No. DE-AC02-06CH11357 and contract no. DE-AC02-76SF00515. T.M. acknowledges financial support from BMBF projects 05K10KT2 and 05K13KT2 as well as DFG BO3169/2-2. This work was supported by the Swedish Research Council, the Knut and Alice Wallenberg Foundation, the European Research Council, the R\"{o}ntgen-Angstr\"{o}m Cluster, ELI Extreme Light Infrastructure Phase 2 (CZ.02.1.01/0.0/0.0/15 008/0000162), ELIBIO (CZ.02.1.01/0.0/0.0/15 003/0000447) from the European Regional Development Fund, Material science and Chalmers Area of Advance. Portions of this research were carried out at Brookhaven National Laboratory, operated under Contract No. DE-SC0012704 from the U.S. Department of Energy (DOE) Office of Science.GF acknowledges the support 
of NKFIH K115504.

     \item[Competing Interests] The authors declare that they have no
    competing financial interests.
     \item[Correspondence] Correspondence and requests for materials
    should be addressed to T.G.~(email: taisgork@slac.stanford.edu).
    \end{addendum}

    \bibliography{Bibfile} 

\begin{thebibliography}{10}
\expandafter\ifx\csname url\endcsname\relax
  \def\url#1{\texttt{#1}}\fi
\expandafter\ifx\csname urlprefix\endcsname\relax\def\urlprefix{URL }\fi
\providecommand{\bibinfo}[2]{#2}
\providecommand{\eprint}[2][]{\url{#2}}

\bibitem{Loh2012}
\bibinfo{author}{Loh, N.~D.} \emph{et~al.}
\newblock \bibinfo{title}{Fractal morphology, imaging and mass spectrometry of
  single aerosol particles in flight}.
\newblock \emph{\bibinfo{journal}{Nature}} \textbf{\bibinfo{volume}{486}},
  \bibinfo{pages}{513--517} (\bibinfo{year}{2012}).

\bibitem{Barke2015}
\bibinfo{author}{Barke, I.} \emph{et~al.}
\newblock \bibinfo{title}{The 3d-architecture of individual free silver
  nanoparticles captured by x-ray scattering}.
\newblock \emph{\bibinfo{journal}{Nature Communications}}
  \textbf{\bibinfo{volume}{6}} (\bibinfo{year}{2015}).

\bibitem{Gomez2014}
\bibinfo{author}{Gomez, L.~F.} \emph{et~al.}
\newblock \bibinfo{title}{Shapes and vorticities of superfluid helium
  nanodroplets}.
\newblock \emph{\bibinfo{journal}{Science}} \textbf{\bibinfo{volume}{345}},
  \bibinfo{pages}{906--909} (\bibinfo{year}{2014}).

\bibitem{VanderSchot2015}
\bibinfo{author}{van~der Schot, G.} \emph{et~al.}
\newblock \bibinfo{title}{{Imaging single cells in a beam of live cyanobacteria
  with an X-ray laser.}}
\newblock \emph{\bibinfo{journal}{Nature Communications}}
  \textbf{\bibinfo{volume}{6}}, \bibinfo{pages}{5704} (\bibinfo{year}{2015}).

\bibitem{Seibert2011}
\bibinfo{author}{Seibert, M.~M.} \emph{et~al.}
\newblock \bibinfo{title}{Single mimivirus particles intercepted and imaged
  with an x-ray laser.}
\newblock \emph{\bibinfo{journal}{Nature}} \textbf{\bibinfo{volume}{470}},
  \bibinfo{pages}{78} (\bibinfo{year}{2011}).

\bibitem{Neutze2000}
\bibinfo{author}{Neutze, R.}, \bibinfo{author}{Wouts, R.},
  \bibinfo{author}{van~der Spoel, D.} \& \bibinfo{author}{Weckert, J.,
  E.and~Hajdu}.
\newblock \bibinfo{title}{Potential for biomolecular imaging with femtosecond
  x-ray pulses}.
\newblock \emph{\bibinfo{journal}{Nature}} \textbf{\bibinfo{volume}{406}},
  \bibinfo{pages}{752} (\bibinfo{year}{2000}).

\bibitem{Chapman2011}
\bibinfo{author}{Chapman, H.~N.} \emph{et~al.}
\newblock \bibinfo{title}{Femtosecond x-ray protein nanocrystallography.}
\newblock \emph{\bibinfo{journal}{Nature}} \textbf{\bibinfo{volume}{470}},
  \bibinfo{pages}{73} (\bibinfo{year}{2011}).

\bibitem{Bostedt2012}
\bibinfo{author}{Bostedt, C.} \emph{et~al.}
\newblock \bibinfo{title}{Ultrafast x-ray scattering from single free
  nanoparticles as a probe for transient states of matter}.
\newblock \emph{\bibinfo{journal}{Physical Review Letters}}
  \textbf{\bibinfo{volume}{108}}, \bibinfo{pages}{093401}
  (\bibinfo{year}{2012}).

\bibitem{Gorkhover2016}
\bibinfo{author}{Gorkhover, T.} \emph{et~al.}
\newblock \bibinfo{title}{{Femtosecond and nanometre visualization of
  structural dynamics in superheated nanoparticles}}.
\newblock \emph{\bibinfo{journal}{Nature Photonics}}
  \textbf{\bibinfo{volume}{10}}, \bibinfo{pages}{93--97}
  (\bibinfo{year}{2016}).

\bibitem{Miao1998}
\bibinfo{author}{Miao, J.}, \bibinfo{author}{Sayre, D.} \&
  \bibinfo{author}{Chapman, H.}
\newblock \bibinfo{title}{Phase retrieval from the magnitude of the fourier
  transforms of nonperiodic objects}.
\newblock \emph{\bibinfo{journal}{JOSA A}} \textbf{\bibinfo{volume}{15}},
  \bibinfo{pages}{1662--1669} (\bibinfo{year}{1998}).

\bibitem{Marchesini2003}
\bibinfo{author}{Marchesini, S.} \emph{et~al.}
\newblock \bibinfo{title}{{X-ray image reconstruction from a diffraction
  pattern alone}}.
\newblock \emph{\bibinfo{journal}{Optics Express}}
  \textbf{\bibinfo{volume}{11}}, \bibinfo{pages}{5} (\bibinfo{year}{2003}).

\bibitem{Chapman2010}
\bibinfo{author}{Chapman, H.~N.} \& \bibinfo{author}{Nugent, K.~A.}
\newblock \bibinfo{title}{Coherent lensless x-ray imaging}.
\newblock \emph{\bibinfo{journal}{Nature Photonics}}
  \textbf{\bibinfo{volume}{4}}, \bibinfo{pages}{833--839}
  (\bibinfo{year}{2010}).

\bibitem{Gabor1948}
\bibinfo{author}{Gabor, D.} \emph{et~al.}
\newblock \bibinfo{title}{A new microscopic principle}.
\newblock \emph{\bibinfo{journal}{Nature}} \textbf{\bibinfo{volume}{161}},
  \bibinfo{pages}{777--778} (\bibinfo{year}{1948}).

\bibitem{Goodman2005}
\bibinfo{author}{Goodman, J.~W.}
\newblock \emph{\bibinfo{title}{Introduction to Fourier optics}}
  (\bibinfo{publisher}{Roberts and Company Publishers}, \bibinfo{year}{2005}).

\bibitem{Eisebitt2004}
\bibinfo{author}{Eisebitt, S.} \emph{et~al.}
\newblock \bibinfo{title}{{Lensless imaging of magnetic nanostructures by
  X-ray}}.
\newblock \emph{\bibinfo{journal}{Nature}} \textbf{\bibinfo{volume}{432}},
  \bibinfo{pages}{885--888} (\bibinfo{year}{2004}).

\bibitem{Schlotter2006}
\bibinfo{author}{Schlotter, W.} \emph{et~al.}
\newblock \bibinfo{title}{{Multiple reference Fourier transform holography with
  soft x rays}}.
\newblock \emph{\bibinfo{journal}{Applied Physics Letters}}
  \textbf{\bibinfo{volume}{89}}, \bibinfo{pages}{2004--2007}
  (\bibinfo{year}{2006}).

\bibitem{Marchesini2008}
\bibinfo{author}{Marchesini, S.} \emph{et~al.}
\newblock \bibinfo{title}{Massively parallel x-ray holography}.
\newblock \emph{\bibinfo{journal}{Nature Photonics}}
  \textbf{\bibinfo{volume}{2}}, \bibinfo{pages}{560--563}
  (\bibinfo{year}{2008}).

\bibitem{Geilhufe2014}
\bibinfo{author}{Geilhufe, J.} \emph{et~al.}
\newblock \bibinfo{title}{{Extracting depth information of 3-dimensional
  structures from a single-view X-ray Fourier-transform hologram}}.
\newblock \emph{\bibinfo{journal}{Optics Express}}
  \textbf{\bibinfo{volume}{22}}, \bibinfo{pages}{24959} (\bibinfo{year}{2014}).

\bibitem{Guehrs2010}
\bibinfo{author}{Guehrs, E.} \emph{et~al.}
\newblock \bibinfo{title}{{Wavefield back-propagation in high-resolution X-ray
  holography with a movable field of view.}}
\newblock \emph{\bibinfo{journal}{Optics Express}}
  \textbf{\bibinfo{volume}{18}}, \bibinfo{pages}{18922--18931}
  (\bibinfo{year}{2010}).

\bibitem{Shintake2008}
\bibinfo{author}{Shintake, T.}
\newblock \bibinfo{title}{{Possibility of single biomolecule imaging with
  coherent amplification of weak scattering x-ray photons}}.
\newblock \emph{\bibinfo{journal}{Physical Review E - Statistical, Nonlinear,
  and Soft Matter Physics}} \textbf{\bibinfo{volume}{78}},
  \bibinfo{pages}{1--9} (\bibinfo{year}{2008}).

\bibitem{Xiao2009}
\bibinfo{author}{Xiao, C.} \emph{et~al.}
\newblock \bibinfo{title}{{Structural studies of the giant Mimivirus}}.
\newblock \emph{\bibinfo{journal}{PLoS Biology}} \textbf{\bibinfo{volume}{7}},
  \bibinfo{pages}{0958--0966} (\bibinfo{year}{2009}).

\bibitem{Ferguson2015}
\bibinfo{author}{Ferguson, K.~R.} \emph{et~al.}
\newblock \bibinfo{title}{The atomic, molecular and optical science instrument
  at the linac coherent light source}.
\newblock \emph{\bibinfo{journal}{Journal of Synchrotron Radiation}}
  \textbf{\bibinfo{volume}{22}}, \bibinfo{pages}{492--497}
  (\bibinfo{year}{2015}).

\bibitem{Ekeberg2015}
\bibinfo{author}{Ekeberg, T.} \emph{et~al.}
\newblock \bibinfo{title}{{Three-Dimensional Reconstruction of the Giant
  Mimivirus Particle with an X-Ray Free-Electron Laser}}.
\newblock \emph{\bibinfo{journal}{Physical Review Letters}}
  \textbf{\bibinfo{volume}{114}}, \bibinfo{pages}{1--6} (\bibinfo{year}{2015}).

\bibitem{Guinier1955}
\bibinfo{author}{Guinier, A.} \& \bibinfo{author}{Fournet, G.}
\newblock \emph{\bibinfo{title}{Small-angle scattering of X-rays}}.
\newblock Structure of matter series (\bibinfo{publisher}{Wiley},
  \bibinfo{year}{1955}).

\bibitem{Bostedt2010}
\bibinfo{author}{Bostedt, C.} \emph{et~al.}
\newblock \bibinfo{title}{Clusters in intense flash pulses: ultrafast
  ionization dynamics and electron emission studied with spectroscopic and
  scattering techniques}.
\newblock \emph{\bibinfo{journal}{J. Phys. B}} \textbf{\bibinfo{volume}{12}},
  \bibinfo{pages}{083004} (\bibinfo{year}{2010}).

\bibitem{Schropp2010}
\bibinfo{author}{Schropp, A.} \& \bibinfo{author}{Schroer, C.~G.}
\newblock \bibinfo{title}{Dose requirements for resolving a given feature in an
  object by coherent x-ray diffraction imaging}.
\newblock \emph{\bibinfo{journal}{New Journal of Physics}}
  \textbf{\bibinfo{volume}{12}}, \bibinfo{pages}{035016}
  (\bibinfo{year}{2010}).

\bibitem{Hantke2014}
\bibinfo{author}{Hantke, M.} \emph{et~al.}
\newblock \bibinfo{title}{{High-throughput imaging of heterogeneous cell
  organelles with an X-ray laser}}.
\newblock \emph{\bibinfo{journal}{Nature Photonics}}
  \textbf{\bibinfo{volume}{8}}, \bibinfo{pages}{943--949}
  (\bibinfo{year}{2014}).

\bibitem{Rupp2016coherent}
\bibinfo{author}{Rupp, D.} \emph{et~al.}
\newblock \bibinfo{title}{Coherent diffractive imaging of single helium
  nanodroplets with a high harmonic generation source}.
\newblock \emph{\bibinfo{journal}{Nature Communications in press, arXiv
  preprint arXiv:1610.05997}}  (\bibinfo{year}{2016}).

\bibitem{Zherebtsov2011}
\bibinfo{author}{Zherebtsov, S.} \emph{et~al.}
\newblock \bibinfo{title}{Controlled near-field enhanced electron acceleration
  from dielectric nanospheres with intense few-cycle laser fields}.
\newblock \emph{\bibinfo{journal}{Nature Physics}}
  \textbf{\bibinfo{volume}{7}}, \bibinfo{pages}{656--662}
  (\bibinfo{year}{2011}).

\bibitem{Gorkhover2012}
\bibinfo{author}{Gorkhover, T.} \emph{et~al.}
\newblock \bibinfo{title}{Nanoplasma dynamics of single large xenon clusters
  irradiated with superintense x-ray pulses from the linac coherent light
  source free-electron laser}.
\newblock \emph{\bibinfo{journal}{Physical Review Letters}}
  \textbf{\bibinfo{volume}{108}}, \bibinfo{pages}{245005}
  (\bibinfo{year}{2012}).

\bibitem{Emma2010}
\bibinfo{author}{Emma, P.} \emph{et~al.}
\newblock \bibinfo{title}{First lasing and operation of an \aa
  ngstrom-wavelength free-electron laser}.
\newblock \emph{\bibinfo{journal}{Nature Photonics}}
  \textbf{\bibinfo{volume}{4}}, \bibinfo{pages}{641} (\bibinfo{year}{2010}).

\bibitem{Rupp2012}
\bibinfo{author}{Rupp, D.} \emph{et~al.}
\newblock \bibinfo{title}{Identification of twinned gas phase clusters by
  single-shot scattering with intense soft x-ray pulses}.
\newblock \emph{\bibinfo{journal}{New Journal of Physics}}
  \textbf{\bibinfo{volume}{14}}, \bibinfo{pages}{055016}
  (\bibinfo{year}{2012}).

\bibitem{Gutt2012}
\bibinfo{author}{Gutt, C.} \emph{et~al.}
\newblock \bibinfo{title}{Single shot spatial and temporal coherence properties
  of the slac linac coherent light source in the hard x-ray regime}.
\newblock \emph{\bibinfo{journal}{Physical Review Letters}}
  \textbf{\bibinfo{volume}{108}}, \bibinfo{pages}{024801}
  (\bibinfo{year}{2012}).

\bibitem{Strueder2010}
\bibinfo{author}{Str\"uder, L.} \emph{et~al.}
\newblock \bibinfo{title}{Large-format, high-speed, x-ray pnccds combined with
  electron and ion imaging spectrometers in a multipurpose chamber for
  experiments at 4th generation light sources}.
\newblock \emph{\bibinfo{journal}{Nucl. Instrum. Meth. A}}
  \textbf{\bibinfo{volume}{614}}, \bibinfo{pages}{483} (\bibinfo{year}{2010}).

\bibitem{Howells2001}
\bibinfo{author}{Howells, M.} \emph{et~al.}
\newblock \bibinfo{title}{{Toward a practical X-ray Fourier holography at high
  resolution}}.
\newblock \emph{\bibinfo{journal}{Nuclear Instruments and Methods in Physics
  Research Section A}} \textbf{\bibinfo{volume}{467-468}},
  \bibinfo{pages}{864--867} (\bibinfo{year}{2001}).

\bibitem{He2004}
\bibinfo{author}{He, H.} \emph{et~al.}
\newblock \bibinfo{title}{{Use of extended and prepared reference objects in
  experimental Fourier transform x-ray holography}}.
\newblock \emph{\bibinfo{journal}{Applied Physics Letters}}
  \textbf{\bibinfo{volume}{85}}, \bibinfo{pages}{2454--2456}
  (\bibinfo{year}{2004}).

\end{thebibliography}
    \bibliographystyle{naturemag}

\end{document}